\documentclass[12pt,a4paper, amssymb, amsmath]{revtex4}
\usepackage{epsfig}
\usepackage{graphicx}
\usepackage{indentfirst}
\usepackage{multirow}
\usepackage{latexsym}

\begin{document}

\title{Electromechanics in MoS$_2$ and WS$_2$: nanotubes vs. monolayers}

\author{Mahdi~Ghorbani-Asl,$^1$ Nourdine~Zibouche,$^{1,2}$ Mohammad~Wahiduzzaman,$^1$\\
Augusto~F.~Oliveira,$^{1,2}$ Agnieszka~Kuc,$^1$ \& Thomas~Heine$^1$\\
$^1$ School of Engineering and Science, Jacobs University Bremen,\\ Campus Ring 1, 28759 Bremen, Germany,\\
$^2$ Scientific Computing \& Modelling NV, Theoretical Chemistry,\\ Vrije Universiteit, De Boelelaan 1083, 1081 HV Amsterdam, The Netherlands}

\begin{abstract}
The transition-metal dichalcogenides (TMD)  MoS$_2$ and WS$_2$ show remarkable electromechanical properties. Strain modifies the direct band gap into an indirect one, and substantial strain even induces an semiconductor-metal transition.
Providing strain through mechanical contacts is difficult for TMD monolayers, but state-of-the-art for TMD nanotubes. We show using density-functional theory that similar electromechanical properties as in monolayer and bulk TMDs are found for large diameter TMD single- (SWNT) and multi-walled nanotubes (MWNTs). The semiconductor-metal transition occurs at elongations of 16~\%. We show that Raman spectroscopy is an excellent tool to determine the strain of the nanotubes and hence monitor the progress of that nanoelectromechanical experiment {\it in situ}. TMD MWNTs show twice the electric conductance compared to SWNTs, and each wall of the MWNTs contributes to the conductance proportional to its diameter.
\end{abstract}
\maketitle

\section*{Introduction}

Transition-metal dichalcogenides (TMDs) have been investigated for the last five decades, mostly for applications in catalysis and as lubricants.\cite{Wilson1969, Mattheis1973a, Kam1982, Tenne1985, Coehoorn1987, Kobayashi1995, Sienicki1996, Gourmelon1997, Wilcoxon1997, Drummond2001}
TMDs form a class of materials with the formula TX$_2$, where T is a transition metal from groups IV--VI (e.g.\ Mo, Ti, Nb), and X is a chalcogen atom (S, Se, or Te).
Bulk TMDs (2$H$) consist of hexagonal tri-atomic layers, X--T--X, with a plane of metal atoms covalently bound to two planes of chalcogen atoms.
Adjacent layers are held together by weak interlayer interactions, what allows exfoliation of individual TX$_2$ slabs.
The electronic properties of bulk TMDs range from metallic to semiconducting, depending on the metal type.

In 2011, TMD materials have gained renewed interest after the successful production of two-dimensional (2D) large-area single layers (1$H$) using liquid exfoliation.\cite{Coleman2011}
Semiconducting TMD monolayers offer properties that are distinct from those of their parental bulk forms.
Such layer-dependent properties have recently attracted great attention for possible applications in nano- and optoelectronics.
Several semiconducting TMDs undergo a transition from the indirect band gap 2$H$ forms to the direct band gap 1$H$-monolayers, as e.g.\ in the case of MoS$_2$, the prototypical TMD material.\cite{Li2007, Splendiani2010, Matte2010, Kuc2011}
Quantum confinement to single layers increases the electronic band gap by about 0.5--1.0~eV, shifting the light absorption from the near-infrared to the visible-light range.
Recently, these novel 2D systems have been used to fabricate first 1$H$-MoS$_2$ field-effect transistors (FETs), logical circuits and amplifiers.\cite{Radisavljevic2011, Radisavljevic2011a, Radisavljevic2012}
Further studies have revealed that TMDs are highly flexible under tensile strain and exhibit excellent pliability,\cite{Bertolazzi2011, Yun2012, Castellanos2012, Johari2012, Zhou2012, Ghorbani2013, Radisavljevic2013} which could be used together with elastic-polymer substrates for development of flexible electronic devices.

Similar to carbon, TMDs form tubular and fullerene-like nanostructures.\cite{Tenne1992, Margulis1993} Though less than their carbon counterparts, in particular MoS$_2$ and WS$_2$ nanoonions and nanotubes have been investigated both theoretically and experimentally.\cite{Seifert2000, Milosevic2007, Tenne2010, Zibouche2012}
Single-walled TMD nanotubes (SWTMD NTs) have interesting electronic properties that depend on their diameter and chirality.
While zigzag ($n$,0) NTs are direct band gap semiconductors resembling 1$H$ forms, the armchair ($n$,$n$) NTs are closer to the 2$H$ structures.\cite{Seifert2000, Zibouche2012}
Increasing the tube diameter, the band gap increases and for diameters larger than 40~\AA\ it approaches the single-layer limit.
TMD NTs have been investigated for their mechanical properties under tensile strain using atomic force microscopy.\cite{Kaplan-Ashiri2004, Kaplan-Ashiri2005, Kaplan-Ashiri2006, Tang2013}
Linear strain-stress relation until fracture suggests plastic deformations, and fracture is directly related to the formation of local defects.\cite{Kaplan-Ashiri2005} While the change of electronic properties under strain has been investigated theoretically for TMD monolayers\cite{Yun2012, Ghorbani2013, Scalise2012}, they remain to be explored for the associated tubular structures. 
As the experimental setup for direct tensile tests of nanotubes is state-of-the-art,\cite{Kaplan-Ashiri2005, Tang2013} the application of tensile stress on 2D TMD systems is rather difficult due to the excellent lubricating properties of these materials. 

\begin{figure}
\includegraphics[scale=0.40,clip]{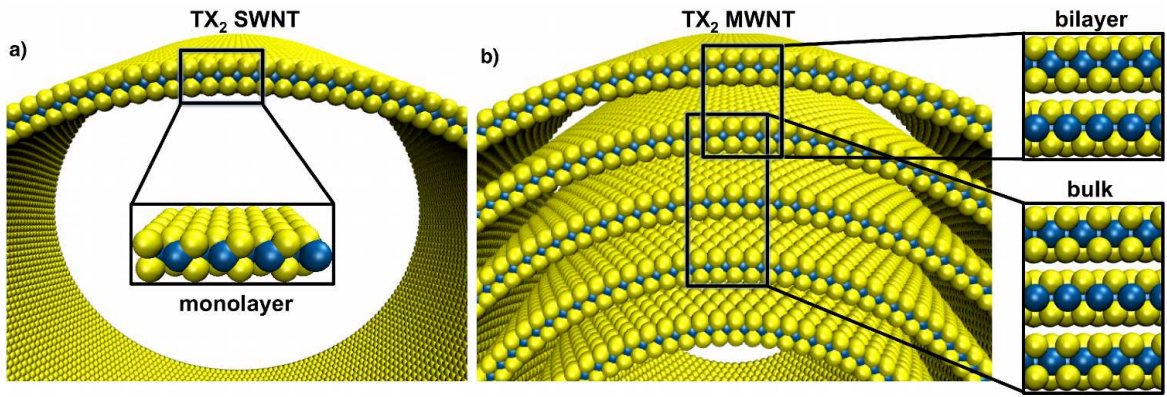}
\caption{\label{fig:1} Structure models of TMD NTs: (a) monolayer represents single-walled nanotubes, (b) bulk and bilayer represent inner and outer walls of multi-walled NTs, respectively. Yellow -- S, blue -- metal (T) atoms.}
\end{figure}

Here, we estimate the electronic, vibrational and electromechanical properties of large-diameter MoS$_2$ and WS$_2$ NTs, using a simple structural model assuming the walls to be planar (see Fig. \ref{fig:1}). We  approximate the single-walled NT (SWNT) by the 1$H$-monolayer, the outer wall of the multi-walled NTs (MWNTs), as well as a hypothetical double-walled NT (DWNT), by the 2$H$ bilayer, and the inner walls of the MWNTs by the 2$H$ bulk structure.
These calculations are relevant for typical experimental MWNT, and the limit for small tubes we have reported earlier.\cite{Zibouche2012} 

We employ density-functional theory (DFT) for all vibrational and electronic structure calculations and the density-functional based tight-binding (DFTB) method for quantum transport calculations (see Methods). The systems are subject to tensile strain up to an elongation of 16~\%, that is before first structural defects occur. We will demonstrate that the individual walls of MWNTs contribute equally, proportional to their diameter, to the electronic transport, and that strain has a homogeneous effect on the structures. We further demonstrate that Raman spectroscopy is an ideal tool to monitor the tensile tests in experiment, as the wavenumber reduces almost linearly by about 3~cm$^{-1}$ per percent of strain for the E$_{2g}$ mode.  

\section*{Results}

\begin{figure}
\includegraphics[scale=0.25,clip]{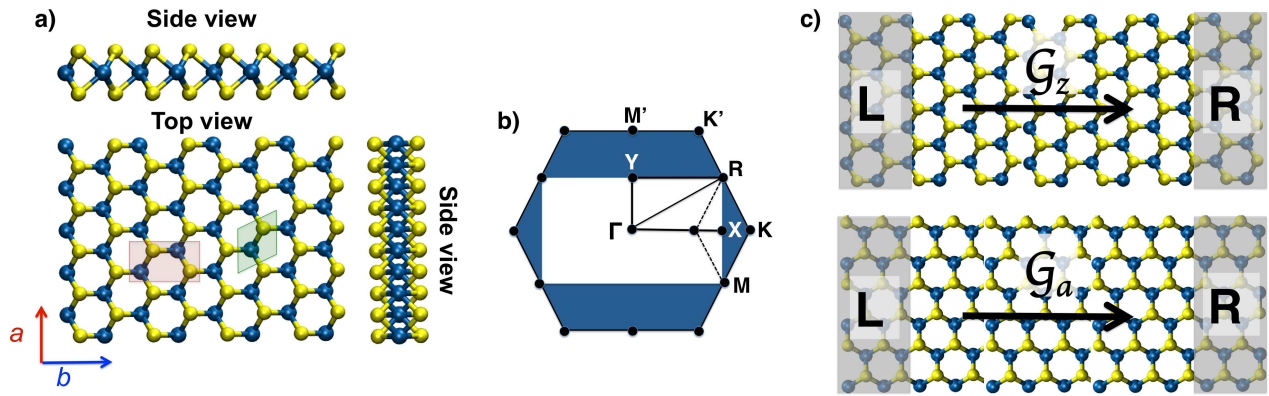}
\caption{\label{fig:BZ} (a) The atomic structure of 2D transition-metal dichalcogenides of TS$_2$ type, T = Mo, W. The $a$ and $b$ lattice vectors are shown. The hexagonal (green) and rectangular (red) unit cells are indicated. (b) The first Brillouin zone of 1$H$- and corresponding rectangular TMD monolayers with the high-symmetry $k$-points. (c) Schematic representation of the direction of the electronic transport, indicated by the quantum conductance $\mathcal{G}$, in the direction normal to the zigzag ($\mathcal{G}_z$) and armchair ($\mathcal{G}_a$) lattice planes. The electrodes, $L$ and $R$, are highlighted with shaded rectangles.}
\end{figure}

In the rectangular representation, the $a$ and $b$ lattice vectors are independent (see Fig.~\ref{fig:BZ}) and refer to tensile strain along armchair and zigzag directions, respectively.
Since we work with rectangular models, the first Brillouin zone (BZ) changes comparing with the hexagonal symmetry (see Fig.~\ref{fig:BZ}).
The BZ of our models is constructed by a proper folding of the hexagonal BZ to the rectangular one,\cite{Jungthawan2011}
where the $K$ point in the hexagonal BZ is folded into the middle of the line that connects it to the $\Gamma$ point.
In the rectangular BZ, this point is located at 2/3 of the symmetry line between $\Gamma$ and $X$ points, and the $X$ and $Y$ points indicate the zone edges.
The $M$ point is located at the same place as $R$ corner point in the rectangular representation.

\begin{figure}
\includegraphics[scale=0.70,clip]{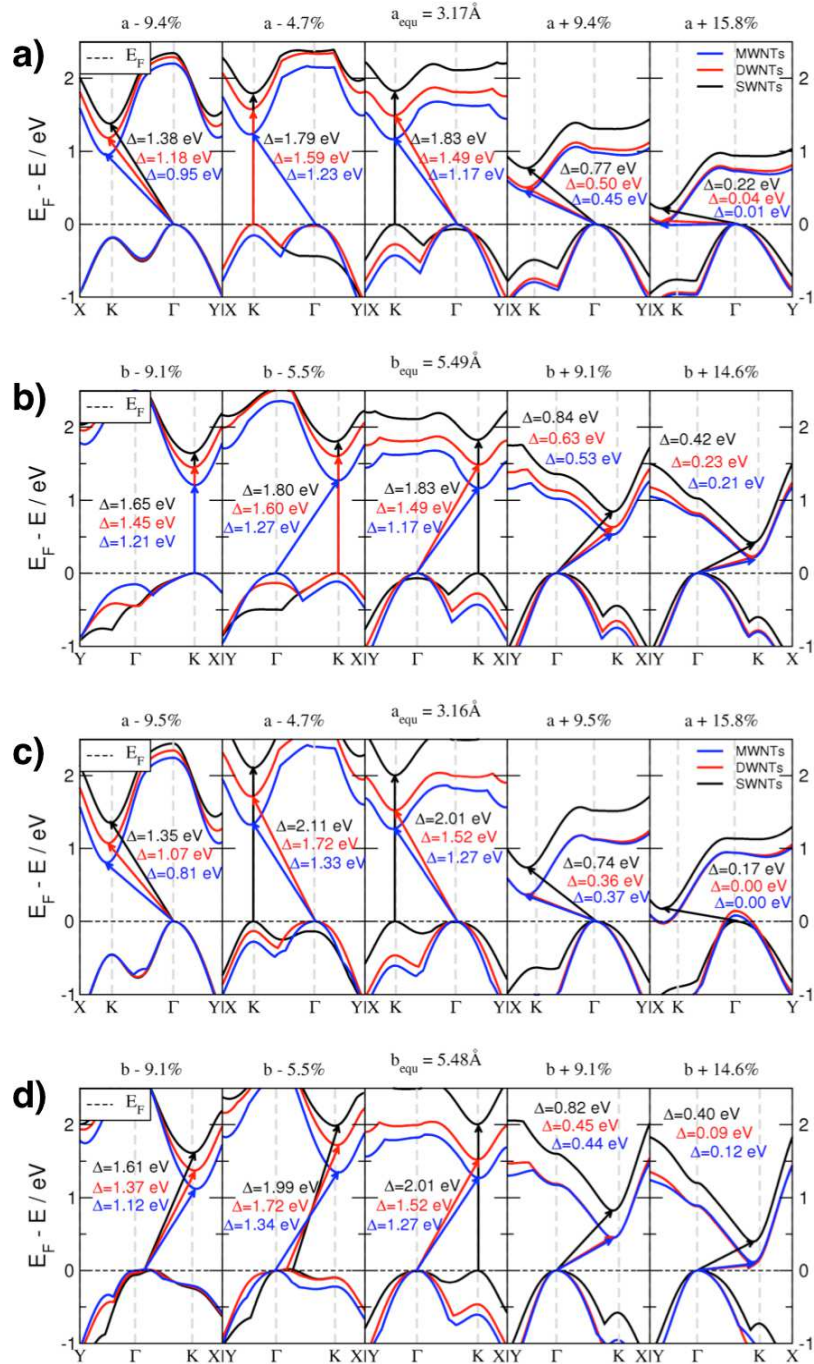}
\caption{\label{fig:BS} Band structures calculated along the armchair (a and c) and zigzag (b and d) directions of MoS$_2$ (a and b) and WS$_2$ (c and d) NTs under tensile strain. Only the valence band maxima and conduction band minima are shown, for clarity. The fundamental band gaps ($\Delta$) are given. The Fermi level is shifted to the top of valence band. MW-, DW-, SWNTs indicate inner walls of MWNTs (corresponding to bulk structures), outer layers of MWNTs and DWNTs (corresponding to bilayers), and SWNTs (corresponding to monolayers), respectively. The amount of applied strain is indicated as percentage relative to the lattice vectors.}
\end{figure}

In order to correctly sample the BZ of NTs, we need to take into account the axial and the circumferential directions.
These correspond to the $k$-paths of $X$--$\Gamma$ and $\Gamma$--$Y$ for the ($n$,$n$) NTs, respectively.
For the ($n$,0) NTs these $k$-paths are reversed.
Since the fundamental band gap is positioned along the $\Gamma$--$K$ line for the 1$H$- and 2$H$-TMDs, $\Delta$ of ($n$,$n$) NTs is obtained by sampling along the tube axis, while for the ($n$,0) NTs we need to sample the circumference.
The resulting band structures of MoS$_2$ and WS$_2$ materials under strain and compression are shown in Fig.~\ref{fig:BS}.

As in the planar structures, $\Delta$ increases with reducing the number of layers, reaching about 2.0~eV for single-layers, while the indirect band gap transforms into a direct one.\cite{Splendiani2010, Kuc2011} Thus, only hypothetical SWNTs, and possibly MWNTs with alternating layer compositions, may show direct band gaps. Slight mechanical deformation of the SWNTs would result in a change of the direct band gap back to the indirect one, located between $\Gamma$ and $K$ high-symmetry points, similarly to the monolayers.\cite{Ghorbani2013} 
A general property is that tensile strain ($\varepsilon$) linearly reduces $\Delta$.
At 9.0~\% elongation, $\Delta$ accounts for less than 50~\%  of its original value. This result is independent of chirality and number of layers, and is thus a global property of MWNTs.
Further stretching results in a semiconductor-metal transition, with the valence band maximum (VBM) crossing the Fermi level at the $\Gamma$ point and the conduction band minimum (CBM) around the $K$ point.
While isotropic stretching of monolayers results in the metallic character of the materials for $\varepsilon\approx11$~\%,\cite{Yun2012, Ghorbani2013} for the tubular systems this transition is observed at much larger elongations of ($\sim$16~\%).
Such large $\varepsilon$ values are at the limit of rupture. In our simulations, the bonding network is still intact at these values. Interestingly, the DW- and MWNTs behave alike under high tensile strain.

\begin{figure}
\includegraphics[scale=0.50,clip]{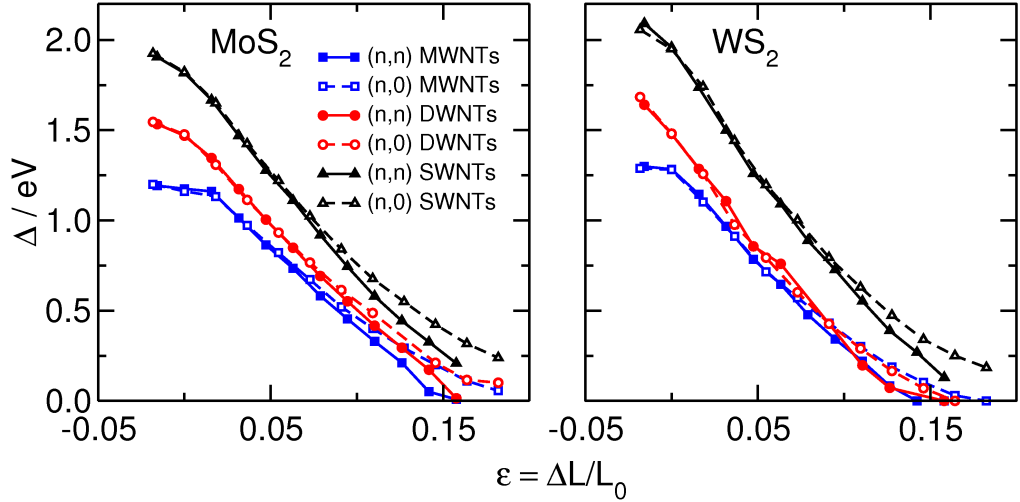}
\caption{\label{fig:gaps} Calculated band gap energies with respect to the applied tensile strain of MoS$_2$ (left) and WS$_2$ (right) NTs.}
\end{figure}

The band gap evolution under tensile strain shows almost linear behavior up to about 12~\% elongation (see Fig.~\ref{fig:gaps}). The slope, however, depends on the chirality, with ($n$,$n$) NTs becoming metallic earlier than the ($n$,0) ones.

\begin{figure}
\includegraphics[scale=0.40,clip]{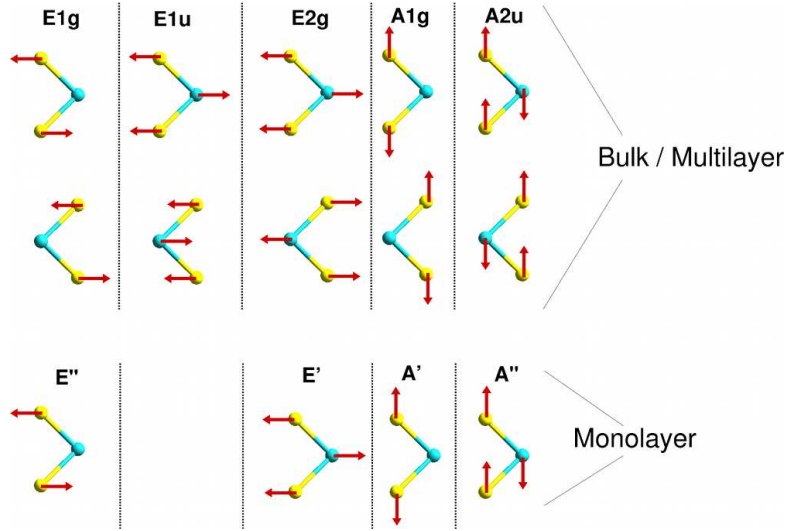}
\caption{\label{fig:Modes} Schematic representation of phonon active modes and their symmetries of bulks and monolayer TS$_2$ (T = Mo, W).}
\end{figure}

The lattice dynamics of 2$H$-MoS$_2$, and more recently also of 1$H$-MoS$_2$, has been studied both experimentally and theoretically.\cite{Lee2010, Ataca2011, Tongay2012, Li2012, Molina2011, Chakraborty2012, Kaasbjerg2012, Rice2013} 
2$H$-MoS$_2$ and 2$H$-WS$_2$ have five active modes, among which $E_{1_u}$ and $A_{1_u}$ are infra-red (IR) active, and $E_{1_g}$, $E_{2_g}$, and $A_{1_g}$ are Raman active (see Fig.~\ref{fig:Modes}).
The $E$ type phonon branches correspond to the in-plane normal modes, while the $A$ type phonons result from the out-of-plane vibrations.
Decreasing the layer thickness to the 1$H$ forms breaks inversion symmetry and thus eliminates the $E_{1_u}$ asymmetric phonons. This results in the four corresponding active modes $E''$, $E'$, $A'$, and $A''$.
The activity of the modes does not change, except for the bulk $E_{2_g}$ mode, which becomes IR and Raman active in the monolayer limit.

We have calculated the phonon dispersion of MoS$_2$ and WS$_2$ in the rectangular unit cell representations (cf. Fig.~\ref{fig:1} c).
Quantum confinement (3D to 2D) causes slight changes in the lattice dynamics (see Tab.~\ref{tab:Phonons_equ}).
For MoS$_2$, the $A_{1_g}$ mode near 409~cm$^{-1}$ becomes softer, transforming into the $A'$ mode near 406~cm$^{-1}$, whereas the $E_{2_g}$ mode near 382~cm$^{-1}$ hardens, transforming into the $E'$ mode near 383~cm$^{-1}$. WS$_2$ systems show very similar behavior (see Tab.~\ref{tab:Phonons_equ}).
\begin{table}
\footnotesize{
\caption{\label{tab:Phonons_equ} Calculated Raman and IR phonon active modes, their symmetry representations, and frequencies (cm$^{-1}$) of MoS$_2$ and WS$_2$ nanotubes modeled as rectangular monolayers (SWNTs), bilayers (DWNTs), and bulk (MWNTs) forms. Available experimental data\cite{Verble1970, Wieting1971, Tongay2012, Gutierrez2013} of corresponding layered materials are given in parenthesis.}
\begin{tabular}{c|c|c|c||c|c|c||c|c|c}
\hline
\multicolumn{4}{c||}{\textbf{Symmetry/Activity}} & \multicolumn{3}{c||}{\textbf{MoS$_2$}} & \multicolumn{3}{c}{\textbf{WS$_2$}} \\
\hline
\multicolumn{2}{c|}{\textbf{MW/DW}} & \multicolumn{2}{c||}{\textbf{SW}} & \textbf{MW} & \textbf{DW} & \textbf{SW} & \textbf{MW} & \textbf{DW} & \textbf{SW} \\
\hline
\textbf{$E_{1_g}$} & R  & \textbf{$E''$} & R       & 285.5 (287) & 285.6 & 285.4               & 283.7              & 283.8  & 283.4              \\
\textbf{$E_{1_u}$} & IR &                        &           & 381.9 (384) & 382.4 &                          & 335.6              & 336.5  &                         \\
\textbf{$E_{2_g}$} & R  & \textbf{$E'$}  & IR+R & 382.1 (383) & 382.8 & 383.3 (384.7) & 336.2 (355.5) & 336.8  & 337.2 (356.0) \\
\textbf{$A_{1_g}$} & R  & \textbf{$A'$}  & R       & 409.4 (409) & 406.9 & 406.3 (406.1) & 414.7 (420.5) & 412.8  & 412.2 (417.5) \\
\textbf{$A_{2_u}$} & IR & \textbf{$A''$} & IR       & 473.4 (470) & 475.0 & 477.2              & 412.6               & 414.9 & 416.6               \\
\hline
\end{tabular}
}
\end{table}

\begin{figure}
\includegraphics[scale=0.12,clip]{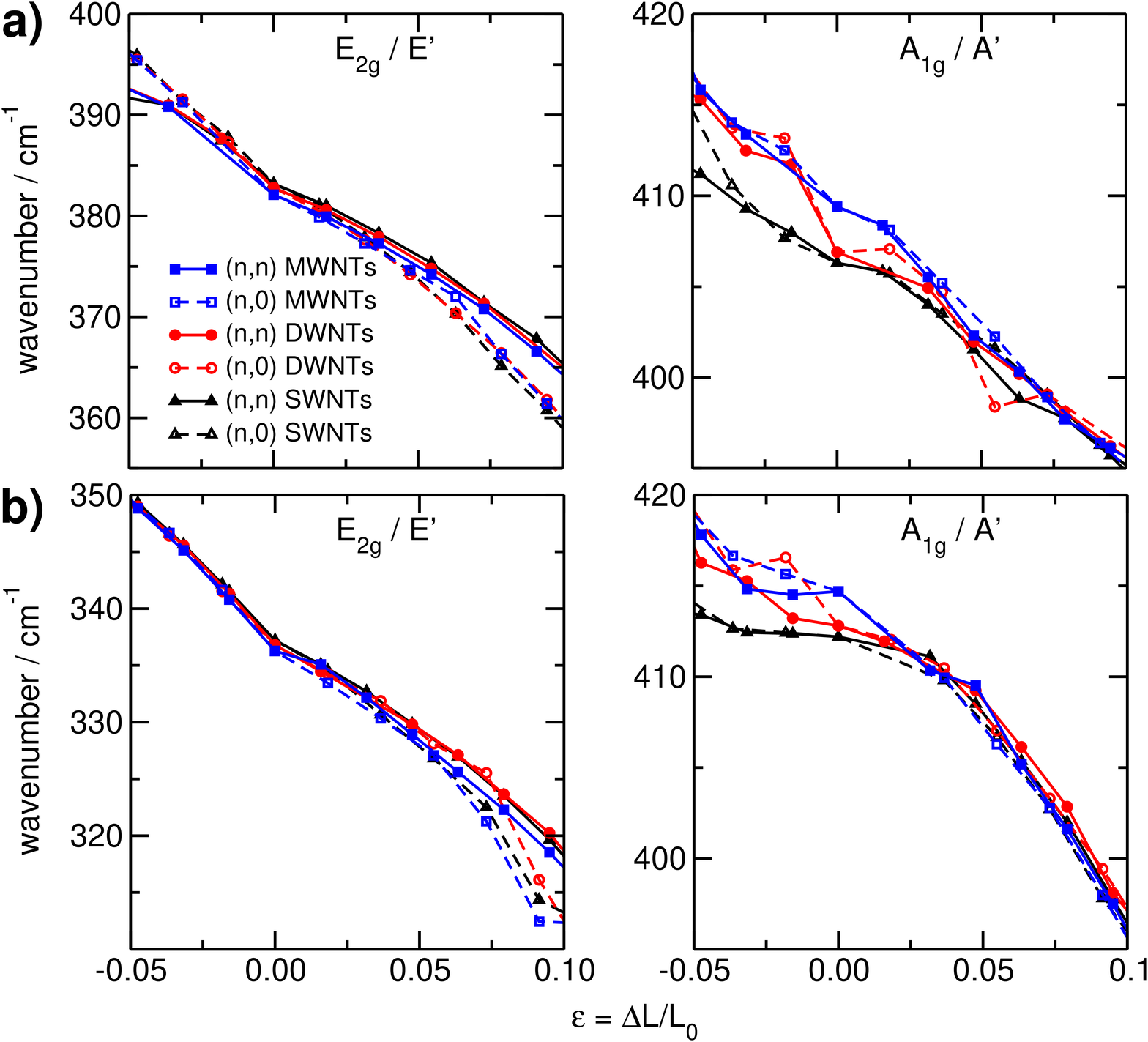}
\caption{\label{fig:Phonons} Calculated phonon $A_{1_g}$ ($E'$) and $E_{2_g}$ ($A'$) active modes of MoS$_2$ (top) and WS$_2$ (bottom) under tensile strain.}
\end{figure}

In contrast to quantum confinement, mechanical deformations strongly alter the lattice dynamics.
For a series of strains we have calculated the Raman-active $E_{2_g}$ ($E'$) and $A_{1_g}$ ($A'$) modes and compared them with those of the equilibrium structures (see Fig.~\ref{fig:Phonons}).
As expected, for both types of active modes, the frequencies reduce with applied strain, whereas they increase under compression. The frequency change is almost linear over a long range of elongations and similar for different chiralities up to a strain of $\sim$8~\%. The slope of this frequency change is about -3 cm$^{-1}$ per percent of strain  for the  $E_{2_g}$ ($E'$) modes and -1 cm$^{-1}$ per percent of strain for the $A_{2_g}$ ($A'$) modes (see Table S1 in the Supporting Information). 

\begin{figure}
\includegraphics[scale=0.55,clip]{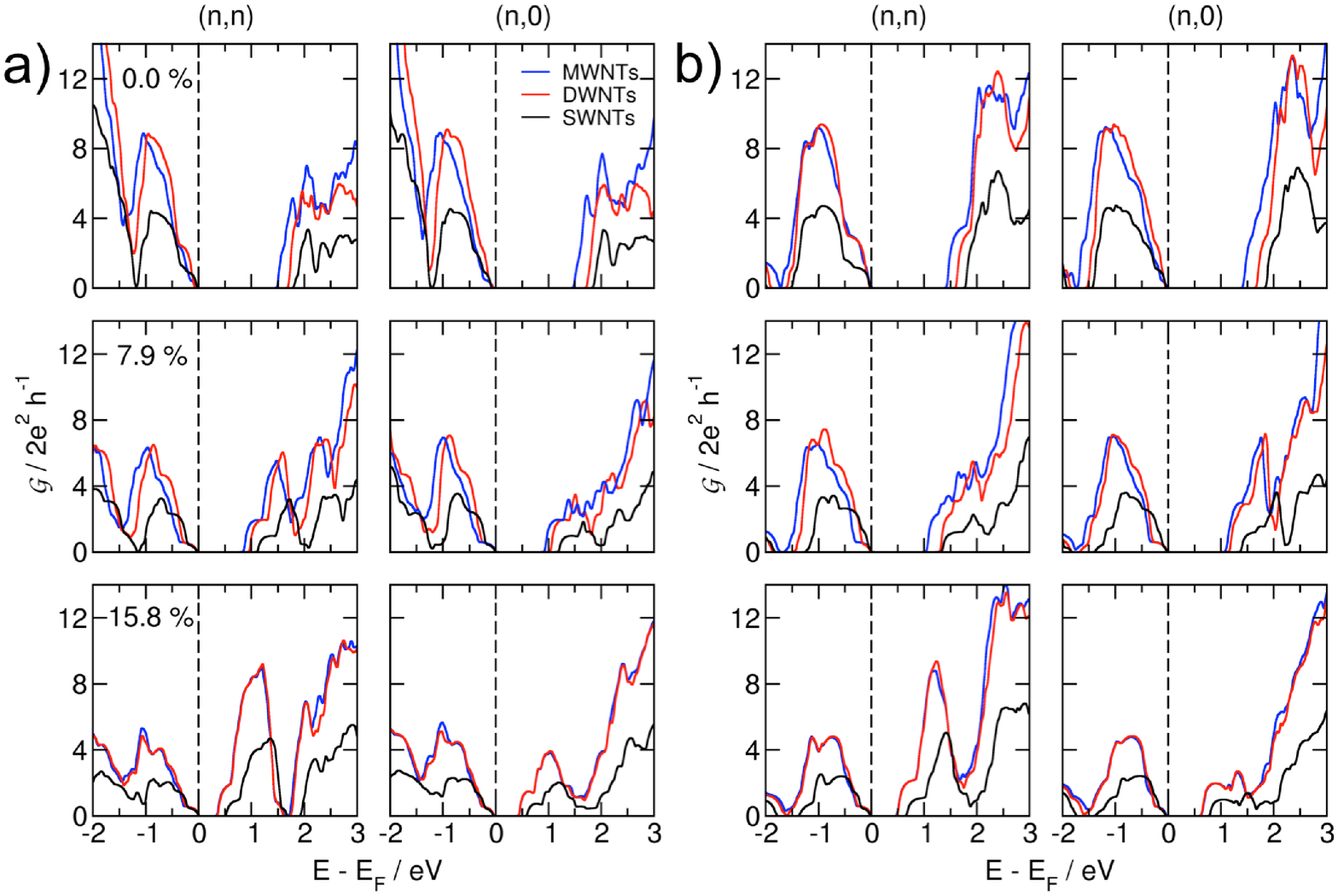}
\caption{\label{fig:Conduct} Intrinsic quantum conductance of MoS$_2$ (a) and WS$_2$ (b) of NTs and layered materials under tensile strain.}
\end{figure}

Figure~\ref{fig:Conduct} shows the intrinsic quantum conductance ($\mathcal{G}$) calculated along the ($n$,$n$) and ($n$,0) TS$_2$ NTs with respect to the applied tensile strain.
As the materials are stretched along the tube axis, $\mathcal{G}$ starts to appear closer to the Fermi level and eventually the transport channel opens.
While for the isotropic stretching of 1$H$-TMDs the transport channels are available already at $\varepsilon$ = 10--11~\%,\cite{Ghorbani2013} for nanotubes these numbers shift to $\sim$15--16~\%.
Note that the quantum transport calculations are carried out using the DFTB method, which tends to overestimate the electronic band gap.

The intensity of the quantum conductance below the Fermi level reduces with applied strain.
The average $\mathcal{G}$ values at $E=$ -1.0~eV for 15.8~\% of elongation are nearly half of those of the equilibrium systems (see Fig.~\ref{fig:Conduct}).
However, for the energies above the Fermi level, chirality matters and $\mathcal{G}$ is higher in the case of ($n$,$n$) tubes.

\section*{Discussion}
For large tubes, quantum confinement will not alter the electronic properties due to the different tube diameters,\cite{Zibouche2012} but it plays a role for the outermost walls of MWNTs. Quantum confinement to the monolayer would be present in SWNTs, and possibly also for recently reported MWNTs with alternating layer compositions that are available for some stoichiometries,\cite{Tang2013} but, so far, not for pure TS$_2$ NTs. 
As for 2D materials, quantum confinement to single-walled tubes would result in direct band-gap semiconductors with $\Delta$ occurring at the $K$ point.\cite{Splendiani2010, Kuc2011}
For smaller tubes, the band gap type depends on the tube chirality, i.e. the ($n$,0) TMD SWNTs are direct-band gap materials, while the ($n$,$n$) ones exhibit indirect $\Delta$ between $K$ and $\Gamma$, which changes slowly to direct-band gap for larger tube diameters.\cite{Zibouche2012, Seifert2000}
This supports the choice of our layered models to represent the large-diameter TMD NTs.
Tensile strain along the tube axis reduces $\Delta$ almost linearly and, for $\varepsilon=$ 16~\%, it closes and a semiconductor-metal transition occurs.

The phonon dispersion is only slightly affected by the position of the individual tube within the MWNTs, as similarly found for the small but apparent quantum confinement effect for layered TMDs. 
Single-walled tubes exhibit slightly softer out-of-plane $A'$ and stronger in-plane $E'$ modes.
Those results are in very good agreement with a large number of available experimental data on layered TMDs, e.g.\ for the in-plane mode it agrees within $\sim$3~cm$^{-1}$ of difference for MoS$_2$ and $\sim$5~cm$^{-1}$ for WS$_2$.\cite{Verble1970, Wieting1971, Tongay2012, Rice2013, Zhang2013, Terrones2013, Gutierrez2013}
The results are likely to be general for the other chalcogenides, for example Tongay et al.\cite{Tongay2012} observed a similarly small increased frequency of the $E_{2_g}$  and decrease of the $A_{1_g}$ modes for MoSe$_2$ layers from Raman measurements.
Those results indicate that the weak interlayer interactions in TX$_2$ materials cannot be associated with the van der Waals interactions only, but most probably with Coulomb electrostatic interactions as well.\cite{Lee2010}

Most theoretical results reported in the literature agree that the quantum confinement effect on the phonon dispersion is rather small. For example,  
Molina-Sanchez et al.\cite{Molina2011} reported the softening of $A_{1_g}$ (from 412.0 to 410.3~cm$^{-1}$) and hardening of $E_{2_g}$ (from 387.8 to 391.7~cm$^{-1}$) modes, based on LDA-DFT calculations.
Ataca et al.\cite{Ataca2011} obtained rather different results on the basis of London dispersion corrected DFT (PW91 functional), and attributed this discrepancy to a rather poor description of the lattice geometry at this level of theory.

The phonon dispersion results for the unstrained structures imply that the thermal properties, such as heat capacity and transport along the tubes, scale linearly with the amount of tube material. They also imply that vibrational spectra of bulk TMD material and MWNT would be very similar, with slight broadening of the signals for the MWNT.

Tensile strain does have a strong effect on the phonon dispersion.
It causes softening of the in-plane $E$ modes by $\sim$3~cm$^{-1}$ per percent of strain, and of the out-of-plane $A$ modes by $\sim$1~cm$^{-1}$ per percent of strain.
After about $\varepsilon=$5~\%, the phonon frequency become dependent on the chirality.
Both, the $A_{1g}$ and $E_{2g}$ modes are Raman active.
Hence, Raman spectroscopy qualifies as an excellent tool to monitor tensile tests of TMDs, both in 2D and in tubular forms.
For 2D materials, strain may be induced by elongation of an appropriate substrate, e.g. by uniform mechanical strain, or by using a material with high thermal expansion coefficient and varying the temperature. For TMD MWNT, tensile tests have been reported by various groups. However, to date, it is not perfectly clear whether inner and outer walls are stretched simultaneously, or rather the outer walls slide on the inner ones. The latter hypothesis would result in a broadening of the Raman signals, while the first one would leave the signal widths rather unaffected. In any case, there would be a shift of the Raman signals that can serve as precise scale for determining the strain.

We have observed exactly the same trends for MoS$_2$ and WS$_2$ nanotubes, both for interlayer interactions and tube stretching. 
Comparing the influence of the two different transition metals, the phonon frequencies of WS$_2$ are generally lower compared to the corresponding MoS$_2$ values, what we rationalize with the larger mass of tungsten atoms and their resulting lower vibrations.
The notable exception is the $A_{1_g}$ mode ($A'$ in 2D), where only the sulfur atoms are vibrating and the frequency is, therefore, affected just by the strength of the T--S covalent bond.\cite{Molina2011} 
Significant lowering of the frequency (by 50--60~cm$^{-1}$) of the $E_{1_u}$, $E_{2_g}$, and $A_{1_u}$ modes in WS$_2$ is due to the fact that these modes include both vibrations of the metal and chalcogen atoms.
The difference between the $E_{2_g}$ ($E'$) and $A_{1_g}$ ($A'$) is now 3-4 times larger than in the case of MoS$_2$.

Our quantum transport calculations aim to describe the intrinsic conductance of the entire tubes along their principal symmetry axis. Conductance normal to the tube axis has not been studied in this work.
Our results show that the transport channels are completely delocalized within the individual walls that compose the MWNT, both for the equilibrium structures and for the strained tubes, even for those strongly strained ones that show zero electronic band gap. Thus, TMD NTs exhibit a truly metallic character if strained by $\sim$16~\%.

It should be noted that band structures, densities of states and phonon dispersion relations have been calculated at the PBE-DFT level, while the quantum conductance calculations employ the DFTB method, an approximation to DFT. While - though somewhat surprisingly - the PBE results give very accurate band gaps for TMD materials,\cite{Kuc2011, Li2013} thorough testing of the DFTB parameters\cite{Ghorbani2013} show an overestimation of the band gap, expressed by a uniform shift of the conduction band to higher levels, while the band forms and thus all other electronic properties are in excellent agreement with DFT. So, the DFTB values for the conduction band should be shifted to lower values, leading to an open transport channel and hence to a truly metallic character in the unbiased $\sim$16\% stretched MWNTs.

The quantum conductance per individual nanotube is about twice higher in MWNTs compared to the corresponding SWNT. Only slight differences in conductivity are observed between the outer and the inner tube walls, provided that they are subject to the same strain. Thus, the entire cross section of the strained or biased MWNTs may serve as conductor if contacted appropriately. 

The quantum conductance results agree very well with our electronic structure calculations (cf. Fig.~\ref{fig:BS}) and with recent experimental data: indeed, the mobility and conductivity of MoS$_2$ thin films increase from mono- to bilayer.\cite{Ghatak2011, Pu2012, Zhang2012, Radisavljevic2013}
The electric double layer transistors (EDLT), fabricated by Zhang and co-workers, showed an electron mobility of 86~cm$^{-2}$ V$^{-1}$ s$^{-1}$ and a high electron density of 10$^{14}$~cm$^{-2}$, which leads to effective electron transport in the EDLT .\cite{Zhang2012}
Kim et al.\cite{Kim2012} have pointed out that a multi-layer MoS$_2$ shows a notably higher current flow in the ballistic regime compared to a single layer device.
This can be attributed to the larger number of transport channels available for the electron flow in the multi-layered MoS$_2$ based junctions.
We estimate that the above results should be transferable to the tubular models, where the MWNTs will reveal better transport properties than the SWNTs.

In conclusion, we have investigated the electronic and lattice response to the tensile strain of TMD nanotubes with different number of layers.
Large-diameter nanotubes can be approximated with layered systems as their properties should be nearly the same at the scale.
Our findings suggest that we can control electronic properties of TMD NTs and layers by an external tensile strain for nanoelectromechanical applications. Raman spectroscopy is an excellent tool to monitor the effect of strain in the samples.

\section*{Methods}
For large-diameter nanotubes, three structural models are needed to describe the individual walls:
A SWNT corresponds to a monolayer, a DWNT to a bilayer. For a MWNT, we distinguish the outermost wall, represented by a bilayer and the inner walls, that are modeled by the bulk material. The innermost layer is not discussed here. In case of inner walls with small diameter we refer to our previous work that discusses the size-dependence of inorganic nanotubes,\cite{Zibouche2012} and for large diameters again the results of the bilayer model would apply. 
All equilibrium structures were fully optimized. For the strained structures, only the coordinates have been optimized, while the unit cell parameters were kept fixed, as done in previous work.\cite{Kuc2011, Ghorbani2013}
The uniaxial tensile strain is applied to the lattice directions $a$ and $b$, resulting in mechanical deformations along the tube axis for the ($n$,$n$) and ($n$,0) nanotubes, respectively (see Fig.~\ref{fig:BZ} a).
The strain is defined as $\varepsilon=\frac{L-L_{0}}{L_{0}}$, where ${L_{0}}$ and ${L}$ are equilibrium and strained lattice values, respectively.

Structural, electronic and vibrational properties have been calculated using DFT and the PBE (Perdew-Burke-Ernzerhof) functional,\cite{PBE} as implemented in the Crystal09 code.\cite{Crystal09}
We employ the computational details that have been validated in our recent studies on TMD structures,\cite{Kuc2011, Li2013} i.e. two-dimensional (single- and bilayers) and three-dimensional periodic boundary conditions (bulk).
For the sulfur atoms, the all-electron 86-311G* basis was chosen, while for the heavier elements the effective core potential (ECP) approach with large cores was employed, accounting for scalar relativistic effects.\cite{Cora1997, Cora1996}
The shrinking factor was set to 9, resulting in 365 $k$ points for bulk structure and 41 $k$ points for the bilayer and monolayer in the irreducible Brillouin zone according to the Monkhorst-Pack sampling.\cite{Monkhorst1976}
Band structures were calculated along the high symmetry points using the following paths \textit{$X$--$\Gamma$--$Y$} and \textit{$Y$--$\Gamma$--$X$} for the ($n$,$n$) and ($n$,0) nanotubes, respectively (see Fig.~\ref{fig:BZ} b).

The coherent electronic transport calculations were carried out using density functional based tight-binding (DFTB)\cite{Seifert1996, Oliveira2009} method in conjunction with the Green's function technique\cite{dicarlo2002, Datta2005} and the Landauer-B\"{u}ttiker approach.
We have already successfully applied this approach to layered TMD materials in their hexagonal representations.\cite{Ghorbani2012, Ghorbani2013}
Transport properties through a material were calculated such that the system is divided into three parts: the semi-infinite left (L) and right (R) leads, and the finite central region (C), also called a scattering region (see Fig.~\ref{fig:BZ} c).
The direction perpendicular to the transport axis is assumed to be infinite by applying periodic boundary conditions.
In this direction, the periodicity is described within the $\Gamma$-point approximation, with a sufficiently large number of unit cells.\cite{Ghorbani2012}
The scattering region was selected to be sufficiently large in order to avoid direct interaction between two semi-infinite leads.
At zero-bias, the conductance ($\mathcal{G}$) is related to the scattering properties of the system by the Fisher-Lee formula:\cite{Fisher1981}
\begin{equation}
\mathcal{G}(E)=\frac{2e^{2}}{h}Tr\left[\hat{G}_{C}^{\dagger}\hat{\,\Gamma}_{R}\hat{\, G}_{C}\hat{\,\Gamma}_{L}\right],
\end{equation}
where $\hat{G}_{C}$ is the Green's function of the scattering part and $\hat{\,\Gamma}_{\alpha}$ are the coupling matrices with $\alpha = L, R$.

The Slater-Koster parameters for DFTB calculations were obtained self-consistently using the PBE density functional with numerical atomic orbitals. The self-consistent optimization has been reported elsewhere\cite{Wahiduzzaman2013} and the BAND software has been employed for this purpose.\cite{ADF, ADF1, ADF2} Details of the performance of the optimized DFTB parameters are given in the Supporting Information.

\section*{Acknowledgements}

This work was supported by the German Research Council (Deutsche Forschungsgemeinschaft, HE 3543/18-1), the European Commission (FP7-PEOPLE-2009-IAPP QUASINANO, GA 251149 and FP7-PEOPLE-2012-ITN MoWSeS, GA 317451). The authors declare that they have no competing financial interests.

\section*{Author contributions}
M.~Ghorbani-Asl, N.~Zibouche, A.~Kuc and T.~Heine generated, analyzed and discussed the results.
M.~Wahiduzzaman and A.~F.~Oliveira generated the DFTB parameters.
T.~Heine conceived this project.
All authors contributed in writing this paper.

\section*{Additional information}

Supplementary information accompanies this paper at http://www.nature.com/scientificreports.


\begin{thebibliography}{10}
\expandafter\ifx\csname url\endcsname\relax
  \def\url#1{\texttt{#1}}\fi
\expandafter\ifx\csname urlprefix\endcsname\relax\def\urlprefix{URL }\fi
\providecommand{\bibinfo}[2]{#2}
\providecommand{\eprint}[2][]{\url{#2}}

\bibitem{Wilson1969}
\bibinfo{author}{Wilson, J.~A.} \& \bibinfo{author}{Yoffe, A.~D.}
\newblock \bibinfo{title}{Transition metal dichalcogenides discussion and
  interpretation of observed optical, electrical and structural properties}.
\newblock \emph{\bibinfo{journal}{Adv. Phys.}} \textbf{\bibinfo{volume}{18}},
  \bibinfo{pages}{193} (\bibinfo{year}{1969}).

\bibitem{Mattheis1973a}
\bibinfo{author}{Mattheis, L.~F.}
\newblock \bibinfo{title}{Energy-bands for 2h-nbse2 and 2h-mos2}.
\newblock \emph{\bibinfo{journal}{Phys. Rev. Lett.}}
  \textbf{\bibinfo{volume}{30}}, \bibinfo{pages}{784--787}
  (\bibinfo{year}{1973}).

\bibitem{Kam1982}
\bibinfo{author}{Kam, K.~K.} \& \bibinfo{author}{Parkinson, B.~A.}
\newblock \bibinfo{title}{Detailed photocurrent spectroscopy of the
  semiconducting group-vi transition-metal dichalcogenides}.
\newblock \emph{\bibinfo{journal}{J. Phys. Chem.}}
  \textbf{\bibinfo{volume}{86}}, \bibinfo{pages}{463} (\bibinfo{year}{1982}).

\bibitem{Tenne1985}
\bibinfo{author}{Tenne, R.} \& \bibinfo{author}{Wold, A.}
\newblock \bibinfo{title}{Passivation of recombination centers in n-wse2 yields
  high-efficiency (greater-than-14-percent) photoelectrochemical cell}.
\newblock \emph{\bibinfo{journal}{Appl. Phys. Lett.}}
  \textbf{\bibinfo{volume}{47}}, \bibinfo{pages}{707} (\bibinfo{year}{1985}).

\bibitem{Coehoorn1987}
\bibinfo{author}{Coehoorn, R.}, \bibinfo{author}{Haas, C.} \&
  \bibinfo{author}{Degroot, R.~A.}
\newblock \bibinfo{title}{Electronic-structure of mose2, mos2, and wse2 .2. the
  nature of the optical band-gaps}.
\newblock \emph{\bibinfo{journal}{Phys. Rev. B}} \textbf{\bibinfo{volume}{35}},
  \bibinfo{pages}{6203} (\bibinfo{year}{1987}).

\bibitem{Kobayashi1995}
\bibinfo{author}{Kobayashi, K.} \& \bibinfo{author}{Yamauchi, J.}
\newblock \bibinfo{title}{Electronic-structure and
  scanning-tunneling-microscopy image of molybdenum dichalcogenide surfaces}.
\newblock \emph{\bibinfo{journal}{Phys. Rev. B}} \textbf{\bibinfo{volume}{51}},
  \bibinfo{pages}{17085} (\bibinfo{year}{1995}).

\bibitem{Sienicki1996}
\bibinfo{author}{Sienicki, W.} \& \bibinfo{author}{Hryniewicz, T.}
\newblock \bibinfo{title}{Tungsten diselenide heterojunction photoelectrodes}.
\newblock \emph{\bibinfo{journal}{Sol. Energ. Mater. Sol. C.}}
  \textbf{\bibinfo{volume}{43}}, \bibinfo{pages}{67} (\bibinfo{year}{1996}).

\bibitem{Gourmelon1997}
\bibinfo{author}{Gourmelon, E.} \emph{et~al.}
\newblock \bibinfo{title}{Ms2 (m=w, mo) photosensitive thin films for solar
  cells}.
\newblock \emph{\bibinfo{journal}{Sol. Energ. Mat. Sol. C.}}
  \textbf{\bibinfo{volume}{46}}, \bibinfo{pages}{115} (\bibinfo{year}{1997}).

\bibitem{Wilcoxon1997}
\bibinfo{author}{Wilcoxon, J.~P.}, \bibinfo{author}{Newcomer, P.~P.} \&
  \bibinfo{author}{Samara, G.~A.}
\newblock \bibinfo{title}{Synthesis and optical properties of mos2 and
  isomorphous nanoclusters in the quantum confinement regime}.
\newblock \emph{\bibinfo{journal}{J. Appl. Phys.}}
  \textbf{\bibinfo{volume}{81}}, \bibinfo{pages}{7934} (\bibinfo{year}{1997}).

\bibitem{Drummond2001}
\bibinfo{author}{Drummond, C.}, \bibinfo{author}{Alcantar, N.},
  \bibinfo{author}{Israelachvili, J.}, \bibinfo{author}{Tenne, R.} \&
  \bibinfo{author}{Golan, Y.}
\newblock \bibinfo{title}{Microtribology and friction-induced material transfer
  in ws2 nanoparticle additives}.
\newblock \emph{\bibinfo{journal}{Adv. Funct. Mater.}}
  \textbf{\bibinfo{volume}{11}}, \bibinfo{pages}{348--354}
  (\bibinfo{year}{2001}).

\bibitem{Coleman2011}
\bibinfo{author}{Coleman, J.~N.} \emph{et~al.}
\newblock \bibinfo{title}{Two-dimensional nanosheets produced by liquid
  exfoliation of layered materials}.
\newblock \emph{\bibinfo{journal}{Science}} \textbf{\bibinfo{volume}{331}},
  \bibinfo{pages}{568} (\bibinfo{year}{2011}).

\bibitem{Li2007}
\bibinfo{author}{Li, T.~S.} \& \bibinfo{author}{Galli, G.~L.}
\newblock \bibinfo{title}{Electronic properties of mos2 nanoparticles}.
\newblock \emph{\bibinfo{journal}{J. Phys. Chem. C}}
  \textbf{\bibinfo{volume}{111}}, \bibinfo{pages}{16192--16196}
  (\bibinfo{year}{2007}).

\bibitem{Splendiani2010}
\bibinfo{author}{Splendiani, A.} \emph{et~al.}
\newblock \bibinfo{title}{Emerging photoluminescence in monolayer mos2}.
\newblock \emph{\bibinfo{journal}{Nano Lett.}} \textbf{\bibinfo{volume}{10}},
  \bibinfo{pages}{1271} (\bibinfo{year}{2010}).

\bibitem{Matte2010}
\bibinfo{author}{Matte, H. S. S.~R.} \emph{et~al.}
\newblock \bibinfo{title}{Mos2 and ws2 analogues of graphene}.
\newblock \emph{\bibinfo{journal}{Angew. Chem. Int. Edit.}}
  \textbf{\bibinfo{volume}{49}}, \bibinfo{pages}{4059} (\bibinfo{year}{2010}).

\bibitem{Kuc2011}
\bibinfo{author}{Kuc, A.}, \bibinfo{author}{Zibouche, N.} \&
  \bibinfo{author}{Heine, T.}
\newblock \bibinfo{title}{{Influence of quantum confinement on the electronic
  structure of the transition metal sulfide TS2}}.
\newblock \emph{\bibinfo{journal}{Phys. Rev. B}} \textbf{\bibinfo{volume}{83}},
  \bibinfo{pages}{245213} (\bibinfo{year}{2011}).

\bibitem{Radisavljevic2011}
\bibinfo{author}{Radisavljevic, B.}, \bibinfo{author}{Radenovic, A.},
  \bibinfo{author}{Brivio, J.}, \bibinfo{author}{Giacometti, V.} \&
  \bibinfo{author}{Kis, A.}
\newblock \bibinfo{title}{Single-layer mos2 transistors}.
\newblock \emph{\bibinfo{journal}{Nat. Nanotechnol.}}
  \textbf{\bibinfo{volume}{6}}, \bibinfo{pages}{147} (\bibinfo{year}{2011}).

\bibitem{Radisavljevic2011a}
\bibinfo{author}{Radisavljevic, B.}, \bibinfo{author}{Whitwick, M.~B.} \&
  \bibinfo{author}{Kis, A.}
\newblock \bibinfo{title}{Integrated circuits and logic operations based on
  single-layer mos2}.
\newblock \emph{\bibinfo{journal}{ACS Nano}} \textbf{\bibinfo{volume}{5}},
  \bibinfo{pages}{9934--9938} (\bibinfo{year}{2011}).

\bibitem{Radisavljevic2012}
\bibinfo{author}{Radisavljevic, B.}, \bibinfo{author}{Whitwick, M.~B.} \&
  \bibinfo{author}{Kis, A.}
\newblock \bibinfo{title}{Small-signal amplifier based on single-layer mos2}.
\newblock \emph{\bibinfo{journal}{Appl. Phys. Lett.}}
  \textbf{\bibinfo{volume}{101}}, \bibinfo{pages}{043103}
  (\bibinfo{year}{2012}).

\bibitem{Bertolazzi2011}
\bibinfo{author}{Bertolazzi, S.}, \bibinfo{author}{Brivio, J.} \&
  \bibinfo{author}{Kis, A.}
\newblock \bibinfo{title}{Stretching and breaking of ultrathin mos2}.
\newblock \emph{\bibinfo{journal}{ACS Nano}} \textbf{\bibinfo{volume}{5}},
  \bibinfo{pages}{9703} (\bibinfo{year}{2011}).

\bibitem{Yun2012}
\bibinfo{author}{Yun, W.~S.}, \bibinfo{author}{Han, S.~W.},
  \bibinfo{author}{Hong, S.~C.}, \bibinfo{author}{Kim, I.~G.} \&
  \bibinfo{author}{Lee, J.~D.}
\newblock \bibinfo{title}{Thickness and strain effects on electronic structures
  of transition metal dichalcogenides: 2h-$m{X}_{2}$ semiconductors ($m$ $=$
  mo, w; $x$ $=$ s, se, te)}.
\newblock \emph{\bibinfo{journal}{Phys. Rev. B}} \textbf{\bibinfo{volume}{85}},
  \bibinfo{pages}{033305} (\bibinfo{year}{2012}).

\bibitem{Castellanos2012}
\bibinfo{author}{Castellanos-Gomez, A.} \emph{et~al.}
\newblock \bibinfo{title}{Elastic properties of freely suspended mos2
  nanosheets}.
\newblock \emph{\bibinfo{journal}{Adv. Mater.}} \textbf{\bibinfo{volume}{24}},
  \bibinfo{pages}{772--775} (\bibinfo{year}{2012}).

\bibitem{Johari2012}
\bibinfo{author}{Johari, P.} \& \bibinfo{author}{Shenoy, V.~B.}
\newblock \bibinfo{title}{Tuning the electronic properties of semiconducting
  transition metal dichalcogenides by applying mechanical strains}.
\newblock \emph{\bibinfo{journal}{ACS Nano}} \textbf{\bibinfo{volume}{6}},
  \bibinfo{pages}{5449} (\bibinfo{year}{2012}).

\bibitem{Zhou2012}
\bibinfo{author}{Zhou, Y.} \emph{et~al.}
\newblock \bibinfo{title}{Tensile strain switched ferromagnetism in layered
  nbs2 and nbse2}.
\newblock \emph{\bibinfo{journal}{ACS Nano}} \textbf{\bibinfo{volume}{6}},
  \bibinfo{pages}{9727--9736} (\bibinfo{year}{2012}).

\bibitem{Ghorbani2013}
\bibinfo{author}{Ghorbani-Asl, M.}, \bibinfo{author}{Borini, S.},
  \bibinfo{author}{Kuc, A.} \& \bibinfo{author}{Heine, T.}
\newblock \emph{\bibinfo{journal}{Phys. Rev. B, accepted}}
  (\bibinfo{year}{2013}).

\bibitem{Radisavljevic2013}
\bibinfo{author}{Radisavljevic, B.} \& \bibinfo{author}{Kis, A.}
\newblock \bibinfo{title}{Mobility engineering and metal-insulator transition
  in monolayer mos2}.
\newblock \emph{\bibinfo{journal}{arXiv:1301.4947 [cond-mat.mes-hall]}}
  (\bibinfo{year}{2013}).

\bibitem{Tenne1992}
\bibinfo{author}{Tenne, R.}, \bibinfo{author}{Margulis, L.},
  \bibinfo{author}{Genut, M.} \& \bibinfo{author}{Hodes, G.}
\newblock \bibinfo{title}{Polyhedral and cylindrical structures of tungsten
  disulfide}.
\newblock \emph{\bibinfo{journal}{Nature}} \textbf{\bibinfo{volume}{360}},
  \bibinfo{pages}{444} (\bibinfo{year}{1992}).

\bibitem{Margulis1993}
\bibinfo{author}{Margulis, L.}, \bibinfo{author}{Salitra, G.},
  \bibinfo{author}{Tenne, R.} \& \bibinfo{author}{Talianker, M.}
\newblock \bibinfo{title}{Nested fullerene-like structures}.
\newblock \emph{\bibinfo{journal}{Nature}} \textbf{\bibinfo{volume}{365}},
  \bibinfo{pages}{113--114} (\bibinfo{year}{1993}).

\bibitem{Seifert2000}
\bibinfo{author}{Seifert, G.}, \bibinfo{author}{Terrones, H.},
  \bibinfo{author}{Terrones, M.}, \bibinfo{author}{Jungnickel, G.} \&
  \bibinfo{author}{Frauenheim, T.}
\newblock \bibinfo{title}{{Structure and electronic properties of MoS2
  nanotubes}}.
\newblock \emph{\bibinfo{journal}{Phys. Rev. Lett.}}
  \textbf{\bibinfo{volume}{85}}, \bibinfo{pages}{146} (\bibinfo{year}{2000}).

\bibitem{Milosevic2007}
\bibinfo{author}{Milosevic, I.} \emph{et~al.}
\newblock \bibinfo{title}{Electronic properties and optical spectra of mos2 and
  ws2 nanotubes}.
\newblock \emph{\bibinfo{journal}{Phys. Rev. B}} \textbf{\bibinfo{volume}{76}}
  (\bibinfo{year}{2007}).

\bibitem{Tenne2010}
\bibinfo{author}{Tenne, R.} \& \bibinfo{author}{Redlich, M.}
\newblock \bibinfo{title}{Recent progress in the research of inorganic
  fullerene-like nanoparticles and inorganic nanotubes}.
\newblock \emph{\bibinfo{journal}{Chem. Soc. Rev.}}
  \textbf{\bibinfo{volume}{39}}, \bibinfo{pages}{1423--1434}
  (\bibinfo{year}{2010}).

\bibitem{Zibouche2012}
\bibinfo{author}{Zibouche, N.}, \bibinfo{author}{Kuc, A.} \&
  \bibinfo{author}{Heine, T.}
\newblock \bibinfo{title}{{From layers to : Transition metal disulfides
  TMS<sub>2</sub>}}.
\newblock \emph{\bibinfo{journal}{Eur. Phys. J. B}}
  \textbf{\bibinfo{volume}{85}}, \bibinfo{pages}{1} (\bibinfo{year}{2012}).

\bibitem{Kaplan-Ashiri2004}
\bibinfo{author}{Kaplan-Ashiri, I.} \emph{et~al.}
\newblock \bibinfo{title}{Mechanical properties of individual ws2 nanotubes}.
\newblock \emph{\bibinfo{journal}{Electronic Properties Of Synthetic
  Nanostructures}} \textbf{\bibinfo{volume}{723}}, \bibinfo{pages}{306--312}
  (\bibinfo{year}{2004}).

\bibitem{Kaplan-Ashiri2005}
\bibinfo{author}{Kaplan-Ashiri, I.} \emph{et~al.}
\newblock \bibinfo{title}{Direct tensile tests of individual ws2 nanotubes}.
\newblock \emph{\bibinfo{journal}{Pricm 5: The Fifth Pacific Rim International
  Conference On Advanced Materials And Processing, Pts 1-5}}
  \textbf{\bibinfo{volume}{475-479}}, \bibinfo{pages}{4097--4102}
  (\bibinfo{year}{2005}).

\bibitem{Kaplan-Ashiri2006}
\bibinfo{author}{Kaplan-Ashiri, I.} \emph{et~al.}
\newblock \bibinfo{title}{On the mechanical behavior of tungsten disulfide
  nanotubes under axial tension and compression}.
\newblock \emph{\bibinfo{journal}{Proc. Natl. Acad. Sci. USA}}
  \textbf{\bibinfo{volume}{103}}, \bibinfo{pages}{523} (\bibinfo{year}{2006}).

\bibitem{Tang2013}
\bibinfo{author}{Tang, D.-M.} \emph{et~al.}
\newblock \bibinfo{title}{Revealing the anomalous tensile properties of ws2
  nanotubes by in situ transmission electron microscopy}.
\newblock \emph{\bibinfo{journal}{Nano Lett.}} \textbf{\bibinfo{volume}{13}},
  \bibinfo{pages}{1034} (\bibinfo{year}{2013}).

\bibitem{Scalise2012}
\bibinfo{author}{Scalise, E.}, \bibinfo{author}{Houssa, M.},
  \bibinfo{author}{Pourtois, G.}, \bibinfo{author}{Afanas'ev, V.} \&
  \bibinfo{author}{Stesmans, A.}
\newblock \bibinfo{title}{Strain-induced semiconductor to metal transition in
  the two-dimensional honeycomb structure of mos2}.
\newblock \emph{\bibinfo{journal}{Nano Res.}} \textbf{\bibinfo{volume}{5}},
  \bibinfo{pages}{43} (\bibinfo{year}{2012}).

\bibitem{Jungthawan2011}
\bibinfo{author}{Jungthawan, S.}, \bibinfo{author}{Limpijumnong, S.} \&
  \bibinfo{author}{Kuo, J.-L.}
\newblock \bibinfo{title}{Electronic structures of graphene/boron nitride sheet
  superlattices}.
\newblock \emph{\bibinfo{journal}{Phys. Rev. B}} \textbf{\bibinfo{volume}{84}},
  \bibinfo{pages}{235424} (\bibinfo{year}{2011}).

\bibitem{Lee2010}
\bibinfo{author}{Lee, C.} \emph{et~al.}
\newblock \bibinfo{title}{Anomalous lattice vibrations of single- and few-layer
  mos2}.
\newblock \emph{\bibinfo{journal}{ACS NANO}} \textbf{\bibinfo{volume}{4}},
  \bibinfo{pages}{2695--2700} (\bibinfo{year}{2010}).

\bibitem{Ataca2011}
\bibinfo{author}{Ataca, C.}, \bibinfo{author}{Topsakal, M.},
  \bibinfo{author}{Akturk, E.} \& \bibinfo{author}{Ciraci, S.}
\newblock \bibinfo{title}{A comparative study of lattice dynamics of three- and
  two-dimensional mos2}.
\newblock \emph{\bibinfo{journal}{J. Phys. Chem. C}}
  \textbf{\bibinfo{volume}{115}}, \bibinfo{pages}{16354}
  (\bibinfo{year}{2011}).

\bibitem{Tongay2012}
\bibinfo{author}{Tongay, S.} \emph{et~al.}
\newblock \bibinfo{title}{Thermally driven crossover from indirect toward
  direct bandgap in 2d semiconductors: Mose2 versus mos2}.
\newblock \emph{\bibinfo{journal}{Nano Lett.}} \textbf{\bibinfo{volume}{12}},
  \bibinfo{pages}{5576} (\bibinfo{year}{2012}).

\bibitem{Li2012}
\bibinfo{author}{Li, T.}
\newblock \bibinfo{title}{Ideal strength and phonon instability in single-layer
  mos2}.
\newblock \emph{\bibinfo{journal}{Phys. Rev. B}} \textbf{\bibinfo{volume}{85}},
  \bibinfo{pages}{235407} (\bibinfo{year}{2012}).

\bibitem{Molina2011}
\bibinfo{author}{Molina-Sanchez, A.} \& \bibinfo{author}{Wirtz, L.}
\newblock \bibinfo{title}{Phonons in single-layer and few-layer mos2 and ws2}.
\newblock \emph{\bibinfo{journal}{Phys. Rev. B}} \textbf{\bibinfo{volume}{84}},
  \bibinfo{pages}{155413} (\bibinfo{year}{2011}).

\bibitem{Chakraborty2012}
\bibinfo{author}{Chakraborty, B.} \emph{et~al.}
\newblock \bibinfo{title}{Symmetry-dependent phonon renormalization in
  monolayer mos${}_{2}$ transistor}.
\newblock \emph{\bibinfo{journal}{Phys. Rev. B}} \textbf{\bibinfo{volume}{85}},
  \bibinfo{pages}{161403} (\bibinfo{year}{2012}).

\bibitem{Kaasbjerg2012}
\bibinfo{author}{Kaasbjerg, K.}, \bibinfo{author}{Thygesen, K.~S.} \&
  \bibinfo{author}{Jacobsen, K.~W.}
\newblock \bibinfo{title}{Phonon-limited mobility in n-type single-layer mos2
  from first principles}.
\newblock \emph{\bibinfo{journal}{Phys. Rev. B}} \textbf{\bibinfo{volume}{85}},
  \bibinfo{pages}{115317} (\bibinfo{year}{2012}).

\bibitem{Rice2013}
\bibinfo{author}{Rice, C.} \emph{et~al.}
\newblock \bibinfo{title}{Raman-scattering measurements and first-principles
  calculations of strain-induced phonon shifts in monolayer mos2}.
\newblock \emph{\bibinfo{journal}{Phys. Rev. B}} \textbf{\bibinfo{volume}{87}},
  \bibinfo{pages}{081307} (\bibinfo{year}{2013}).

\bibitem{Verble1970}
\bibinfo{author}{Verble, J.~L.} \& \bibinfo{author}{Wieting, T.~J.}
\newblock \bibinfo{title}{Lattice mode degeneracy in mos2 and other layer
  compounds}.
\newblock \emph{\bibinfo{journal}{Phys. Rev. Lett.}}
  \textbf{\bibinfo{volume}{25}}, \bibinfo{pages}{362} (\bibinfo{year}{1970}).

\bibitem{Wieting1971}
\bibinfo{author}{Wieting, T.~J.} \& \bibinfo{author}{Verble, J.~L.}
\newblock \bibinfo{title}{Infrared and raman studies of long-wavelength optical
  phonons in hexagonal mos2}.
\newblock \emph{\bibinfo{journal}{Phys. Rev. B}} \textbf{\bibinfo{volume}{3}},
  \bibinfo{pages}{4286} (\bibinfo{year}{1971}).

\bibitem{Zhang2013}
\bibinfo{author}{Zhang, X.} \emph{et~al.}
\newblock \bibinfo{title}{Raman spectroscopy of shear and layer breathing modes
  in multilayer mos2}.
\newblock \emph{\bibinfo{journal}{Phys. Rev. B}} \textbf{\bibinfo{volume}{87}}
  (\bibinfo{year}{2013}).

\bibitem{Terrones2013}
\bibinfo{author}{Terrones, H.}, \bibinfo{author}{Lopez-Urias, F.} \&
  \bibinfo{author}{Terrones, M.}
\newblock \bibinfo{title}{Novel hetero-layered materials with tunable direct
  band gaps by sandwiching different metal disulfides and diselenides}.
\newblock \emph{\bibinfo{journal}{Sci. Rep.}} \textbf{\bibinfo{volume}{3}},
  \bibinfo{pages}{1549} (\bibinfo{year}{2013}).

\bibitem{Gutierrez2013}
\bibinfo{author}{Gutierrez, H.~R.} \emph{et~al.}
\newblock \bibinfo{title}{Extraordinary room-temperature photoluminescence in
  triangular ws2 monolayers}.
\newblock \emph{\bibinfo{journal}{Nano Lett.}}
  \textbf{\bibinfo{volume}{available online, DOI: 10.1021/nl3026357}}.

\bibitem{Li2013}
\bibinfo{author}{Li, W.}, \bibinfo{author}{Walther, C. F.~J.},
  \bibinfo{author}{Kuc, A.} \& \bibinfo{author}{Heine, T.}
\newblock \bibinfo{title}{Density functional theory and beyond for band-gap
  screening: Performance for transition-metal oxides and dichalcogenides}.
\newblock \emph{\bibinfo{journal}{J. Chem. Theory Comput.}}
  \textbf{\bibinfo{volume}{available online, DOI: 10.1021/ct400235w}}
  (\bibinfo{year}{2013}).

\bibitem{Ghatak2011}
\bibinfo{author}{Ghatak, S.}, \bibinfo{author}{Pal, A.~N.} \&
  \bibinfo{author}{Ghosh, A.}
\newblock \bibinfo{title}{Nature of electronic states in atomically thin mos2
  field-effect transistors}.
\newblock \emph{\bibinfo{journal}{ACS Nano}} \textbf{\bibinfo{volume}{5}},
  \bibinfo{pages}{7707} (\bibinfo{year}{2011}).

\bibitem{Pu2012}
\bibinfo{author}{Pu, J.} \emph{et~al.}
\newblock \bibinfo{title}{Highly flexible mos2 thin-film transistors with ion
  gel dielectrics}.
\newblock \emph{\bibinfo{journal}{{Nano Lett.}}} \textbf{\bibinfo{volume}{12}},
  \bibinfo{pages}{4013} (\bibinfo{year}{2012}).

\bibitem{Zhang2012}
\bibinfo{author}{Zhang, Y.}, \bibinfo{author}{Ye, J.},
  \bibinfo{author}{Matsuhashi, Y.} \& \bibinfo{author}{Iwasa, Y.}
\newblock \bibinfo{title}{Ambipolar mos2 thin flake transistors}.
\newblock \emph{\bibinfo{journal}{Nano Lett.}} \textbf{\bibinfo{volume}{12}},
  \bibinfo{pages}{1136} (\bibinfo{year}{2012}).

\bibitem{Kim2012}
\bibinfo{author}{Kim, S.} \emph{et~al.}
\newblock \bibinfo{title}{High-mobility and low-power thin-film transistors
  based on multilayer mos2 crystals}.
\newblock \emph{\bibinfo{journal}{Nat. Commun.}} \textbf{\bibinfo{volume}{3}},
  \bibinfo{pages}{1011} (\bibinfo{year}{2012}).

\bibitem{PBE}
\bibinfo{author}{Perdew, J.~P.}, \bibinfo{author}{Burke, K.} \&
  \bibinfo{author}{Ernzerhof, M.}
\newblock \bibinfo{title}{Generalized gradient approximation made simple}.
\newblock \emph{\bibinfo{journal}{Phys. Rev. Lett.}}
  \textbf{\bibinfo{volume}{77}}, \bibinfo{pages}{3865} (\bibinfo{year}{1996}).

\bibitem{Crystal09}
\bibinfo{author}{Dovesi, R.} \emph{et~al.}
\newblock \bibinfo{title}{Crystal09 user's manual. university of torino:
  Torino} (\bibinfo{year}{2009}).

\bibitem{Cora1997}
\bibinfo{author}{Cora, F.}, \bibinfo{author}{Patel, A.},
  \bibinfo{author}{Harrison, N.~M.}, \bibinfo{author}{Roetti, C.} \&
  \bibinfo{author}{Catlow, C. R.~A.}
\newblock \bibinfo{title}{An ab initio hartree-fock study of alpha-moo3}.
\newblock \emph{\bibinfo{journal}{J. Mater. Chem.}}
  \textbf{\bibinfo{volume}{7}}, \bibinfo{pages}{959} (\bibinfo{year}{1997}).

\bibitem{Cora1996}
\bibinfo{author}{Cora, F.}, \bibinfo{author}{Patel, A.},
  \bibinfo{author}{Harrison, N.~M.}, \bibinfo{author}{Dovesi, R.} \&
  \bibinfo{author}{Catlow, C. R.~A.}
\newblock \bibinfo{title}{An ab initio hartree-fock study of the cubic and
  tetragonal phases of bulk tungsten trioxide}.
\newblock \emph{\bibinfo{journal}{J. Am. Chem. Soc.}}
  \textbf{\bibinfo{volume}{118}}, \bibinfo{pages}{12174}
  (\bibinfo{year}{1996}).

\bibitem{Monkhorst1976}
\bibinfo{author}{Monkhorst, H.~J.} \& \bibinfo{author}{Pack, J.~D.}
\newblock \bibinfo{title}{Special points for brillouin-zone integrations}.
\newblock \emph{\bibinfo{journal}{Phys. Rev. B}} \textbf{\bibinfo{volume}{13}},
  \bibinfo{pages}{5188} (\bibinfo{year}{1976}).

\bibitem{Seifert1996}
\bibinfo{author}{Seifert, G.}, \bibinfo{author}{Porezag, D.} \&
  \bibinfo{author}{Frauenheim, T.}
\newblock \emph{\bibinfo{journal}{Int. J. Quantum Chem.}}
  \textbf{\bibinfo{volume}{58}}, \bibinfo{pages}{185} (\bibinfo{year}{1996}).

\bibitem{Oliveira2009}
\bibinfo{author}{Oliveira, A.~F.}, \bibinfo{author}{Seifert, G.},
  \bibinfo{author}{Heine, T.} \& \bibinfo{author}{Duarte, H.~A.}
\newblock \bibinfo{title}{Density-functional based tight-binding: an
  approximate dft method}.
\newblock \emph{\bibinfo{journal}{J. Br. Chem. Soc.}}
  \textbf{\bibinfo{volume}{20}}, \bibinfo{pages}{1193--1205}
  (\bibinfo{year}{2009}).

\bibitem{dicarlo2002}
\bibinfo{author}{Di~Carlo, A.} \emph{et~al.}
\newblock \bibinfo{title}{{Molecular Devices Simulations Based on Density
  Functional Tight-Binding}}.
\newblock \emph{\bibinfo{journal}{J. Comp. Electron.}} \bibinfo{pages}{109}
  (\bibinfo{year}{2002}).

\bibitem{Datta2005}
\bibinfo{author}{Datta, S.}
\newblock \emph{\bibinfo{title}{Quantum Transport: Atom to Transistor}}
  (\bibinfo{publisher}{Cambridge University Press}, \bibinfo{address}{Cambridge
  and New York}, \bibinfo{year}{2005}), \bibinfo{edition}{2} edn.

\bibitem{Ghorbani2012}
\bibinfo{author}{Ghorbani-Asl, M.}, \bibinfo{author}{Juarez-Mosqueda, R.},
  \bibinfo{author}{Kuc, A.} \& \bibinfo{author}{Heine, T.}
\newblock \bibinfo{title}{Efficient quantum simulations of metals within the
  gamma-point approximation: Application to carbon and inorganic 1d and 2d
  materials}.
\newblock \emph{\bibinfo{journal}{J. Chem. Theory Comput.}}
  \textbf{\bibinfo{volume}{8}}, \bibinfo{pages}{2888} (\bibinfo{year}{2012}).

\bibitem{Fisher1981}
\bibinfo{author}{Fisher, D.~S.} \& \bibinfo{author}{Lee, P.~A.}
\newblock \bibinfo{title}{Relation between conductivity and transmission
  matrix}.
\newblock \emph{\bibinfo{journal}{Phys. Rev. B}} \textbf{\bibinfo{volume}{23}},
  \bibinfo{pages}{6851} (\bibinfo{year}{1981}).

\bibitem{Wahiduzzaman2013}
\bibinfo{author}{Wahiduzzaman, M.} \emph{et~al.}
\newblock \bibinfo{title}{Dftb parameters for the periodic table: Part 1,
  electronic structure}.
\newblock \emph{\bibinfo{journal}{Submitted to J. Chem. Theory Comput.}}
  (\bibinfo{year}{2013}).

\bibitem{ADF}
\bibinfo{author}{Philipsen, P. H.~T.} \emph{et~al.}
\newblock \bibinfo{title}{Band2012}.
\newblock \bibinfo{howpublished}{SCM, Theoretical Chemistry, Vrije
  Universiteit, Amsterdam, The Netherlands, http://www.scm.com}
  (\bibinfo{year}{2012}).

\bibitem{ADF1}
\bibinfo{author}{Wiesenekker, G.} \& \bibinfo{author}{Baerends, E.~J.}
\newblock \bibinfo{title}{Quadratic integration over the 3-dimensional
  brillouin-zone}.
\newblock \emph{\bibinfo{journal}{J. Phys.: Condens. Matter}}
  \textbf{\bibinfo{volume}{3}}, \bibinfo{pages}{6721--6742}
  (\bibinfo{year}{1991}).

\bibitem{ADF2}
\bibinfo{author}{Velde, G.~T.} \& \bibinfo{author}{Baerends, E.~J.}
\newblock \bibinfo{title}{Precise density-functional method for periodic
  structures}.
\newblock \emph{\bibinfo{journal}{Phys. Rev. B}} \textbf{\bibinfo{volume}{44}},
  \bibinfo{pages}{7888--7903} (\bibinfo{year}{1991}).

\end{thebibliography}
\end{document}